# Shubnikov-de Haas effect in low electron density SrTiO$_3$: Conduction band edge of SrTiO$_3$ redux

S. James Allen[1], Bharat Jalan[2,*], SungBin Lee[1], Daniel G. Ouellette[1], Guru Khalsa[3],

Jan Jaroszynski[4], Susanne Stemmer[2], and Allan H. MacDonald[3]

[1] *Department of Physics, UC Santa Barbara, Santa Barbara, CA 93106-9530, USA*

[2] *Materials Department, University of California, Santa Barbara, California 93106-5050, USA*

[3] *Department of Physics, The University of Texas at Austin, Austin, Texas, 78712, USA*

[4] *National High Magnetic Field Laboratory, Florida State University, Tallahassee, FL 32310, USA*

**Abstract**

The Shubnikov-de Haas effect is used to explore the conduction band edge of high mobility SrTiO$_3$ films doped with La. The results largely confirm the earlier measurements by Uwe *et al.* [Jap. J. Appl. Phys. **24**, Suppl. **24-2,** 335 (1985)]. The band edge dispersion differs significantly from the predictions of *ab initio* electronic structure theory.

PACS number(s): 71.18.+y,71.38.-k, 71.70.Fk

## I. Introduction

Strontium titanate is a perovskite oxide that is prominently featured in the emerging arena of oxide electronics.[1,2,3] Heterostructures formed between $SrTiO_3$ and other oxides[4] exhibit an interfacial, two dimensional electron gas (2DEG) that can be controlled by applied electric fields[5], and for which relatively high 2D densities on the order of $3\times10^{14}$ cm$^{-2}$ can be achieved[6]. Furthermore, superconductivity[7,8], ordered magnetic ground states[9,10,11,12] and the Kondo effect[13] are observed. In most cases, transport occurs on the $SrTiO_3$-side of the interface. A quantitative description of the conduction band states in bulk $SrTiO_3$ is essential[14] to model and theoretically understand the properties of confined 2DEGs.

The conduction band of bulk $SrTiO_3$, a wide gap band insulator, has been the subject of experimental and theoretical attention for about 50 years.[15,16,17,18,19,20] While early models featured a conduction band edge similar to Si with valleys near the $X$-point, subsequent experiments[21] and modeling placed the conduction band minimum at $\Gamma$, the zone center. Uwe *et al.*[22,23] used the Shubnikov-de Haas effect and Raman scattering to determine the spin-orbit interaction strength and sign, the band edge splitting caused by the low temperature tetragonal distortion, and details of the Fermi surface. Most recently, quantum oscillations of the Nernst effect have been documented in very low doped $SrTiO_3$ and the authors conclude that at sufficiently low doping the conduction electrons occupy a single "barely anisotropic" Fermi surface but details about the shape are not pursued.[24]

The low energy conduction bands can be characterized by three $t_{2g}$ Luttinger parameters, the spin-orbit interaction, and tetragonal distortion energies. More recently Uwe et al.'s results were questioned by ARPES experiments[25]. There has been no consensus on the relative strength of spin-orbit and tetragonal strain parameters responsible for band-splitting at the band edge.

Accurate knowledge of the bulk bands is essential if progress is to be made toward understanding of 2DEGs in SrTiO$_3$.

Here we use the earlier results from Uwe et al. [22,23] as a guide for Shubnikov de-Haas oscillation experiments in high mobility, lightly La-doped SrTiO$_3$ films. Using the parameters determined by Uwe et al. [22,23], Fig. 1 shows the dispersion along [001], and the Fermi surfaces and energy for an electron density of $7.5\times10^{17}$cm$^{-3}$. The model dispersion and Fermi surfaces shown here include the effect of a relatively large spin-orbit interaction and discussed further in our data analysis that follows. At the lowest doping concentration a single closed Fermi surface is expected, while higher concentrations cause the occupation of a higher band, which is split-off at low temperature by the tetragonal distortion of the unit cell. Its Fermi surface provides a tight constraint on the strain-induced splitting. At this concentration two bands are expected to be occupied. Our Shubnikov-de Haas experiments are carried out on samples that range from doping (or electron concentrations) that are sufficiently low that only one Fermi surface is filled, to higher concentrations in which two Fermi surfaces are filled, as depicted in Fig. 1. At the outset we note that although the results reported here differ quantitatively from Uwe et al. [22,23], they substantially support their interpretation, despite the fact that their experiments were carried out at substantially higher doping concentrations, requiring them to extract extremal Fermi surface cross sections that were interconnected by magnetic breakdown.

The results reported here measure the Luttinger parameters[26] for the $t_{2g}$ conduction band minimum. We are able to successfully interpret our experiments by assuming that the spin-orbit energy (measured by Uwe et al. to be ~17 meV [23]) is much larger than the Fermi energy at the electron density of the samples investigated here. The measured Luttinger parameters differ substantially from recent band structure models opening the possibility that the ab initio calculations are not accurate, or the band edge mass is substantially enhanced by electron phonon coupling[27], or a combination of both.

## II. Experimental

SrTiO$_3$ films doped with La were grown on (001) SrTiO$_3$ substrates by molecular beam epitaxy (MBE), as described elsewhere[28,29]. These films exhibit the high mobility needed to observe the Shubnikov-de Haas effect and to explore the conduction band edge. Low temperature (1.8 K) Hall carrier density (calculated as $n = 1/(t \cdot e \cdot R_H)$, where $t$ is the film thickness, $R_H$ the Hall coefficient, and $e$ the elementary charge) and mobilities varied from $3.6 \times 10^{17}$ cm$^{-3}$ (mobility 37,000 cm$^2$V$^{-1}$s$^{-1}$) to $12 \times 10^{17}$ cm$^{-3}$ (mobility 33,000 cm$^2$V$^{-1}$s$^{-1}$). The thickness of the epitaxial layers varied from 800 nm to 1200 nm. While a total of five epitaxial layers were investigated, a complete set of data was taken and analyzed for the two samples referred to in the following as samples 1 and 2, see Table I. Swept field magnetotransport experiments were carried out at the National High Magnetic Field Lab at temperatures down to 0.4 K and magnetic fields (B) to 31 T as a function of angle between the magnetic field and sample normal.

## III. Results and Discussion

Results for sample 1 with the magnetic field aligned along [111] are shown in Fig. 2. The oscillatory features (relative resistance maxima) are indexed and plotted as a function of 1/B in Fig. 3; they display a straight line corresponding to an extremal Fermi surface cross section of ~ 17.5 T. The extremal area, $S_F$, is related to the slope by $S_F = \dfrac{\partial n}{\partial(1/B)} \cdot 4\pi^2 \dfrac{e}{h}$. Figure 2 shows that the quantum limit is reached around this field and the oscillations begin to show a doubling of the resistance maxima as spin split Fermi surfaces are resolved.

Quantum oscillations for sample 2 with the magnetic field aligned along [001], the surface normal, are shown in Fig. 4. At low fields a weak, low frequency oscillation is detected and assigned to the Fermi surface for electrons in the band split-off from the conduction band minimum by the tetragonal strain. As in sample 1, at the highest fields the resistance maxima split, due to the spin split Fermi surfaces. Indices for the relative resistance maxima for the two sets of oscillations are plotted as a function of 1/B in Fig. 5. The corresponding extremal areas for the two Fermi surfaces for sample 2 along [001] are 55.2 and 9.55 T, respectively.

At 105 K, $SrTiO_3$ undergoes a phase transformation from cubic to a tetragonal phase. The tetragonal [001] direction (*c*-axis) can then select three different directions - normal to the sample surface or in two orthogonal directions in the plane of the film, corresponding to three orientation variants or domains. In this case the experimental results for any given orientation of the magnetic field could display several different periods of oscillation each corresponding to a particular domain. Each sample was measured along the three principal directions, [110], [111] and [001] defined such that [001] is the surface normal. Figure 6 shows the measured extremal area for the larger Fermi surface in sample 2 for these orientations. Also shown is the extremal areas calculated using Uwe *et al.*'s parameters and assuming that the tetragonal axis is normal to the surface (90° or [001]) in the figure. There is a numerical discrepancy, but the angle dependence is similar. More importantly, if we calculate the orientation dependence for a domain with the tetragonal axis in the plane, a <010> direction, we find a qualitative difference in the angular dependence. From this we conclude that we are observing quantum oscillations only from domains with the tetragonal c-axis normal to the surface, or, alternatively, that the sample is a single domain, with tetragonal c-axis normal to the sample surface.

Shubnikov-de Haas oscillations were measured for these two samples, each in the three aforementioned orientations. This information is sufficient to determine Fermi surface shapes and consequently the low energy band parameters, subject to an energy scale factor. To determine the energy scale factor, the temperature dependence of a set of quantum oscillations was measured and an effective mass ($m^*$) extracted for that extremal cross section. This is sufficient to establish an energy scale by relating the measured mass to the rate of change of the extremal cross section with energy, $\frac{\partial S_F}{\partial E} = m^* \cdot \frac{2\pi}{\hbar^2}$. Figure 7 shows the temperature dependence of the Shubnikov-de Haas oscillations for sample 2 with the magnetic field oriented along [110]. Also shown is a model calculation[30] that fits the measurements with an effective mass of $1.41 m_e$, for this particular extremal orbit.

## IV Analysis

The experimental data are fit to Fermi surfaces that are described by a band edge effective mass Hamiltonian subject to a large spin-orbit interaction. The key assumption is that the spin-orbit interaction is much larger than the Fermi energies at the doping levels in the samples used here and much larger than the splitting induced by the tetragonal distortion. Then we follow the model of Khalsa and MacDonald.[31]

The band edge effective mass Hamiltonian in the $t_{2g}$ basis, $\{|yz,\sigma\rangle, |zx,\sigma\rangle, |xy,\sigma\rangle\}$, is expressed as:

$$H_{\sigma,\vec{k}} = \begin{bmatrix} \frac{1}{2}(\gamma_1 - 4\gamma_2)k_x^2 + \frac{1}{2}(\gamma_1 + 2\gamma_2)(k_z^2 + k_y^2) + be & 3\gamma_3 k_x k_y & 3\gamma_3 k_x k_z \\ 3\gamma_3 k_x k_y & \frac{1}{2}(\gamma_1 - 4\gamma_2)k_y^2 + \frac{1}{2}(\gamma_1 + 2\gamma_2)(k_z^2 + k_x^2) + be & 3\gamma_3 k_z k_y \\ 3\gamma_3 k_x k_z & 3\gamma_3 k_y k_z & \frac{1}{2}(\gamma_1 - 4\gamma_2)k_z^2 + \frac{1}{2}(\gamma_1 + 2\gamma_2)(k_x^2 + k_y^2) - 2be \end{bmatrix}$$

(1),

independent of spin, $\sigma$, and where $k^2 = k_x^2 + k_y^2 + k_z^2$, and $\gamma_1$, $\gamma_2$ and $\gamma_3$ are closely analogous to the Luttinger[32] parameters commonly used to describe the valence band structure in elemental semiconductors. The effect of the tetragonal strain[33] is parameterized by $be$, following Uwe et al.'s notation where $e$ is the tetragonal strain and $b$ the deformation potential. In Eq. (1) $k$ is dimensionless and equal to *1* at the zone boundary X point, $\pi/a$.

The diagonal components of (1) can be related to an anisotropic effective mass for each of the three $t_{2g}$ states. If we define a heavy effective mass, $m_h$, and relatively lighter transverse effective mass $m_t$ as follows $m_h = \frac{\hbar^2}{m_e}\left(\frac{\pi}{a}\right)^2 \frac{1}{(\gamma_1 - 4\gamma_2)}$ and $m_t = \frac{\hbar^2}{m_e}\left(\frac{\pi}{a}\right)^2 \frac{1}{(\gamma_1 + 2\gamma_2)}$, then the anisotropic mass for each of the three $t_{2g}$ states can be expressed as follows. For $|yz,\sigma\rangle$ we have $m_x^{zy} = m_h$ and $m_y^{zy} = m_z^{zy} = m_t$, for $|xz,\sigma\rangle$ $m_y^{xz} = m_h$ and $m_x^{xz} = m_z^{xz} = m_t$, and for $|xy,\sigma\rangle$ $m_z^{xy} = m_h$, and $m_y^{xy} = m_x^{xy} = m_t$. $\hbar$ and $m_e$ are Planck's constant and the free electron mass

We take the spin-orbit interaction, $\Delta_{SO}$, to be significantly larger than the components of $H_{\sigma,\vec{k}}$. That is to say "band edge" in this analysis implies states with energy much smaller than $\Delta_{SO}$. The total Hamiltonian, including spin-orbit coupling, is:

$$H_{total} = \begin{bmatrix} H_{\uparrow,\vec{k}} & 0 \\ 0 & H_{\downarrow,\vec{k}} \end{bmatrix} + \frac{\Delta_{SO}}{3} \begin{bmatrix} 1 & i & 0 & 0 & 0 & -1 \\ -i & 1 & 0 & 0 & 0 & i \\ 0 & 0 & 1 & 1 & -i & 0 \\ 0 & 0 & 1 & 1 & -i & 0 \\ 0 & 0 & i & i & 1 & 0 \\ -1 & -i & 0 & 0 & 0 & 1 \end{bmatrix}, \begin{Bmatrix} |yz,\uparrow\rangle \\ |zx,\uparrow\rangle \\ |xy,\uparrow\rangle \\ |yz,\downarrow\rangle \\ |zx,\downarrow\rangle \\ |xy,\downarrow\rangle \end{Bmatrix}. \quad (2)$$

To characterize the lowest lying excitations, we transform the total Hamiltonian (2) by forcing the diagonalization of the spin-orbit part. In this limit, the total Hamiltonian $H_{total}$ describes the dispersion of the spin-orbit split off state and the lower band edge states. If we ignore the off-diagonal terms which couple the lower band edge states and the spin-orbit split off part separated by $\Delta_{SO}$, we recover the dispersion of the two fold degenerate, spin-orbit split-off states, $E_{SO} = \Delta_{SO} + \frac{1}{2}\gamma_1 k^2$, and a Hamiltonian that describes the dispersion of the remaining lowest four conduction band states, which participate in the quantum oscillations:

$$\frac{1}{2}\gamma_1 k^2 + \begin{bmatrix} \frac{1}{2}\gamma_2(k_x^2 + k_y^2 - 2k_z^2) - be & \frac{\sqrt{3}}{2}\gamma_2(k_x^2 - k_y^2) + i\gamma_3\sqrt{3}k_x k_y & \sqrt{3}\gamma_3 k_z(k_x + ik_y) & 0 \\ \frac{\sqrt{3}}{2}\gamma_2(k_x^2 - k_y^2) - i\gamma_3\sqrt{3}k_x k_y & -\frac{1}{2}\gamma_2(k_x^2 + k_y^2 - 2k_z^2) + be & 0 & \sqrt{3}\gamma_3 k_z(k_x + ik_y) \\ \sqrt{3}\gamma_3 k_z(k_x - ik_y) & 0 & -\frac{1}{2}\gamma_2(k_x^2 + k_y^2 - 2k_z^2) + be & -\frac{\sqrt{3}}{2}\gamma_2(k_x^2 - k_y^2) - i\gamma_3\sqrt{3}k_x k_y \\ 0 & \sqrt{3}\gamma_3 k_z(k_x - ik_y) & -\frac{\sqrt{3}}{2}\gamma_2(k_x^2 - k_y^2) + i\gamma_3\sqrt{3}k_x k_y & \frac{1}{2}\gamma_2(k_x^2 + k_y^2 - 2k_z^2) - be \end{bmatrix}$$

(3)

The eigenvalues of (3) describe two bands, each two fold degenerate, given by the following:

$$E_\pm = \tfrac{1}{2}\gamma_1 k^2 \pm \left[\gamma_2^2 k^4 - 3(\gamma_2^2 - \gamma_3^2)(k_x^2 k_y^2 + k_x^2 k_z^2 + k_y^2 k_z^2) + \gamma_2 be(2k_z^2 - k_x^2 - k_y^2) + (be)^2\right]^{1/2} \quad (4)$$

This is identical to the expression used by Uwe et al.[22] but expressed in terms of the Luttinger parameters. We note that if we restrict our measurements to energies much less than the spin-orbit splitting, the results are not influenced by the strength of the spin orbit interaction and we determine the parameters for $H_{\sigma,\vec{k}}$ in Eq. (1).

As is the case for the valence band of elemental and compound semiconductors, the complex dispersion relation will lead to a complex spin-Landau spectrum especially at high magnetic fields. Indeed, at the highest fields the quantum oscillations begin to resolve spin dependent Fermi surfaces.

The extremal areas measured at the prescribed angles for the two samples were simultaneously fit, using a non-linear algorithm, to the dispersion relation (4) by adjusting the following dimensionless parameters: $\gamma_2/\gamma_1$, $\gamma_3/\gamma_1$, $E_{F,1}/\gamma_1$, $E_{F,2}/\gamma_1$, and $be/\gamma_1$. These parameters, as ratios, determine the size and shape of the measured Fermi surfaces but not the energy scale; the size and shape of the various extremal areas are independent of $\gamma_1$. By adjusting $\gamma_1$, we can fit the mass determined by the temperature dependence of the Shubnikov-de Haas oscillations. The fitting parameters are then expressed in appropriate energy units.

Table II shows the parameters that were determined from these fits. Satisfactory agreement between the measured and calculated angular dependence using the parameters in Table II, is shown in Fig. 8. Not shown in Fig. 8 is the agreement achieved by the fit for the narrow waist of the Fermi surface that originates from strain split off band; the measured 9.55 Tesla compares well with a fit value of 9.47 Tesla. A more complete low energy model that includes the spin-orbit split off band may increase the Luttinger parameters by ~ 10%.

Apart from ~ 10% differences, the results essentially agree with the earlier results of Uwe et al.[22] The discrepancies may reflect the fact that the earlier work by them was carried out at higher electron densities and was extracted by disentangling orbits that suffered magnetic breakdown. The discrepancy between the Shubnikov-de Haas and Hall densities in the two samples used in this analysis may be due to the moderate complexity of the Fermi surface in this density range.

Various band structure calculations predict Luttinger parameters that are substantially larger than those measured by Shubnikov-de Haas oscillations shown in Table II [20,34]. In particular we compare with a recent *ab initio* band structure calculation by Janotti et al.[34] and conclude that the SdH mass is ~ 2 times heavier than predicted. We can ascribe this discrepancy to strong electron phonon coupling[35] only if we accept at face value the *ab initio* band structure calculations. A phonon enhancement of ~ 2-3 was implied in recent infrared measurements of the extended Drude response.[36]

We also note that the parameters imply a vanishingly small dispersion, $\gamma_1 - 4\gamma_2 \approx 0$, and as a result, a very large heavy mass, $m_h$. As pointed out by Janotti et al.[34], the spin orbit coupling admixes the three $t_{2g}$ states and the dispersion at the band edge (See Fig. 1.) does not show the extreme anisotropy displayed in equation (1) and measured here. However, models of electric subbands in SrTiO$_3$ with extremely high electron densities involve large $\vec{k}$ vectors due to relatively large in plane Fermi energies and tight quantum confinement. Then, models of these quantum confined states are best developed with the extreme anisotropy of equation (1), introducing the spin orbit interaction after electric surface quantization.

**Summary**

Low temperature Shubnikov-de Haas effect measured on high mobility but low density La doped SrTiO$_3$ samples was used to determine the tetragonally induced band edge spitting and low energy Luttinger band edge parameters. The work substantially agrees with Uwe et al.[22]

The band edge Luttinger parameters differ substantially from those predicted by *ab initio* calculations and require an examination of the calculated band structure and/or mass enhancement by electron phonon coupling. Similar experiments in higher carrier density samples could potentially provide an independent measurement of the spin-orbit interaction strength.

**Acknowledgements:** This work benefitted from useful conversations with Anderson Janotti and Chris Van der Walle, and figure formatting by Pouya Moetakef. Work at UC Santa Barbara was supported by the MRSEC Program of the National Science Foundation (Award No. DMR 1121053). Work at the University of Texas was supported by the National Science Foundation under grants DGE-0549417 and DMR-1122603 and by the Welch Foundation under grant TBF1473. The National High Magnetic Field Laboratory is supported by the NSF under DMR-0654118 and the State of Florida.

[36] J. L. M. van Mechelen, D. van der Marel, C. Grimaldi, A. B. Kuzmenko, N. P. Armitage, N. Reyren, H. Hagemann, and I.I. Mazin, Phys. Rev. Lett. **100**, 226403 (2008).

# Figure Captions

**Figure 1 (color online):** Dispersion at the conduction band edge using parameters from Uwe et al.[22] in a model that includes relatively strong spin-orbit interaction. Fermi surfaces correspond to an electron density of $7.5\times10^{17}$ cm$^{-3}$, and $k$ is expressed in units of $\pi/a$ where $a$ is the length of the cubic SrTiO$_3$ unit cell. $k_Z$ is directed along the tetragonal c-axis.

**Figure 2 (color online):** Shubnikov-de Haas oscillations with the magnetic field aligned along [111] for sample 1. (a) The quantum oscillations are exhausted at the quantum limit, ~ 15 Tesla. (b) Features persist down to 2 Tesla.

**Figure 3 (color online):** Sample 1. The relative maxima in Fig. 2, plotted vs. 1/B. Near the quantum limit a splitting appears. The slope of the line corresponds to an extremal area of 17.5 Tesla.

**Figure 4 (color online):** Sample 2. Shubnikov-de Haas oscillations with the magnetic field aligned along [001]. (a) Spin splitting is apparent at the highest field. At low fields a weak, low frequency oscillation is observed (arrows) and assigned to the strain induced split-off band.

**Figure 5 (color online):** Sample 2. Indexed relative maxima for two sets of quantum oscillations in Fig. 4, corresponding to extremal cross sections of 55.2 and 9.55 Tesla.

**Figure 6 (color online):** Sample 2. Angle dependence of measured extremal area, solid squares. A comparison with parameters from Uwe et al.[22] is also shown (open and closed circles, respectively). Open circles assume that the tetragonal *c*-axis is normal to the surface, closed circles assume the tetragonal c-axis is in the plane of the sample.

**Figure 7 (color online):** Temperature dependence of the Shubnikov-de Haas oscillations for sample 2. (a) Measured with the magnetic field along the [110] direction. (b) Model calculation with an effective mass of 1.41 $m_e$.

**Figure 8 (color online):** Fermi surface area versus angle, measured (solid) and fit (open), using the parameters in the text.

Table I: Samples investigated in this study.

| Sample number | Internal sample reference no. | Layer thickness (nm) | 1.8 K Hall electron density ($cm^{-3}$) | 1.8 K Hall mobility ($cm^2V^{-1}s^{-1}$) | SdH electron density ($cm^{-3}$) |
|---|---|---|---|---|---|
| 1 | STO-216 | 1280 | $3.6\times10^{17}$ | 37,000 | $4.2\times10^{17}$ |
| 2 | STO-181 | 800 | $12\times10^{17}$ | 33,000 | $18.2\times10^{17}$ |

Table II: Conduction band parameters determined in this study and comparison with the literature.

| Parameter | Experiment (this study) | Experiment (Uwe et al.[22,23]) | Band structure calculation (Janotti et al.[34]) |
|---|---|---|---|
| $\gamma_1$ | 4.0 ($\pm$0.04) eV | 3.5 eV | 8.81 eV |
| $\gamma_2$ | 0.98 ($\pm$0.02) eV | 0.88 eV | 1.92 eV |
| $\gamma_1+2\gamma_2$ | 6.0 eV | 5.26 eV | 12.65 eV |
| $\gamma_1-4\gamma_2$ | 0.0 ($\pm$0.1) | 0.0 | 1.15 eV |
| $\gamma_3$ | 0.0 ($\pm$0.02) eV | 0.13 eV | 0.78 eV |
| $2be$ | -2.2 meV | -1.5 meV | -2.268 meV |

**Figures**

Fig. 1

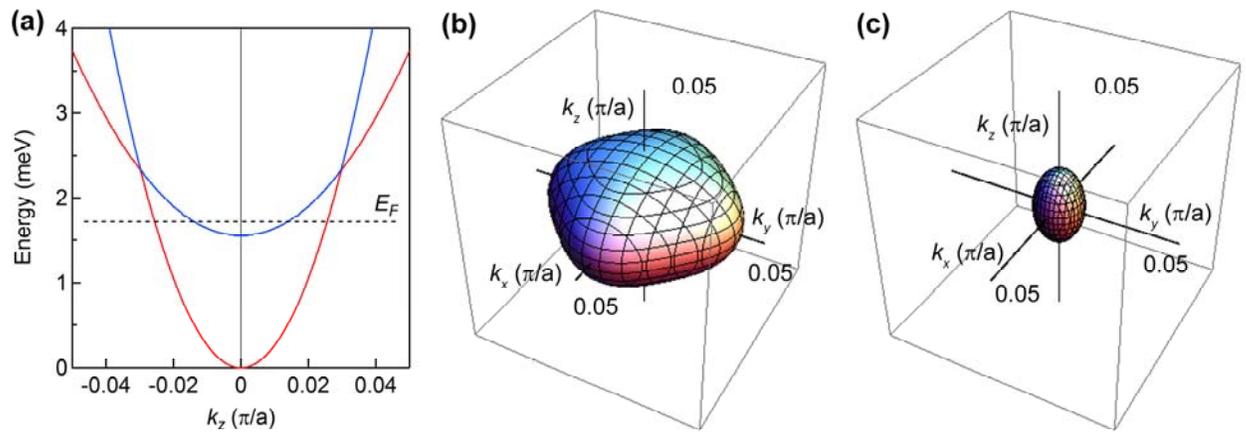

Fig. 2

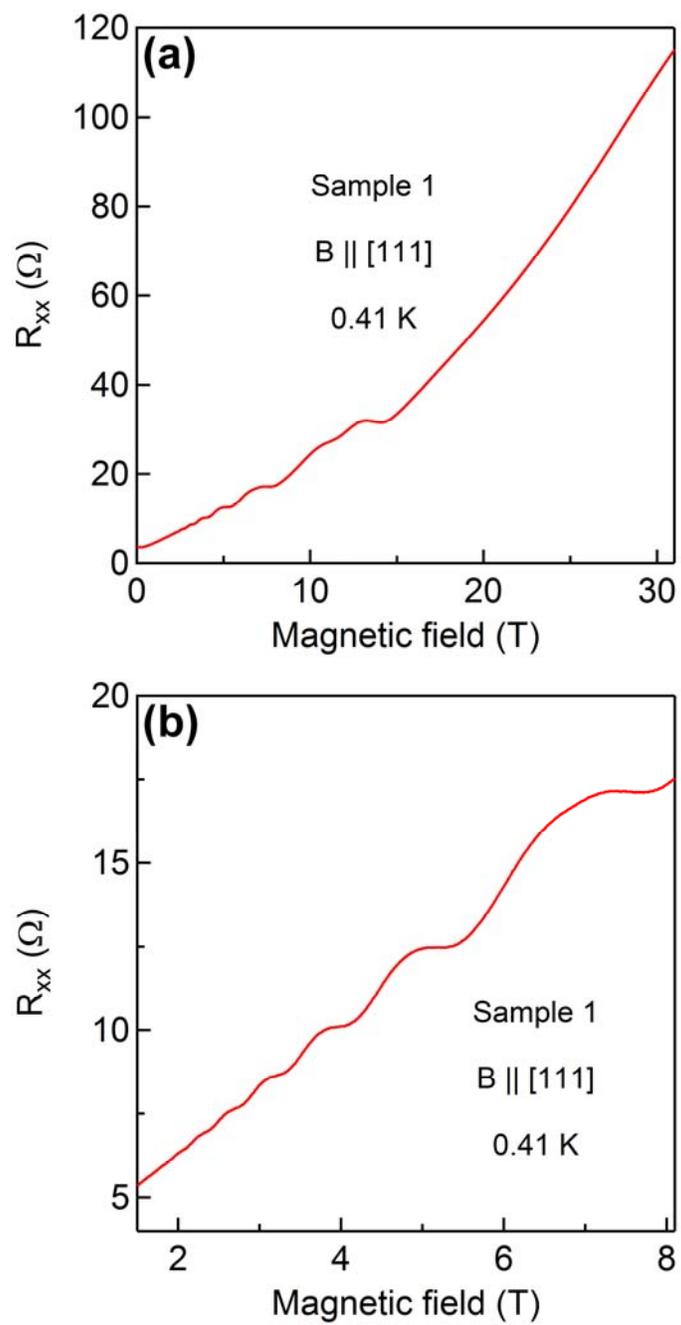

Fig. 3

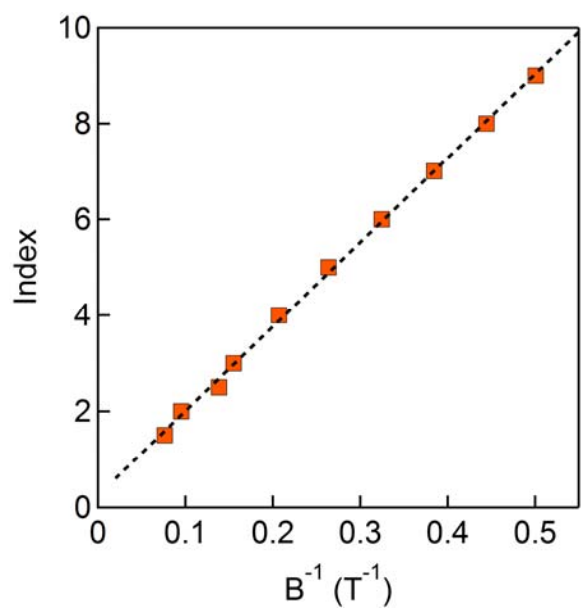

Fig. 4

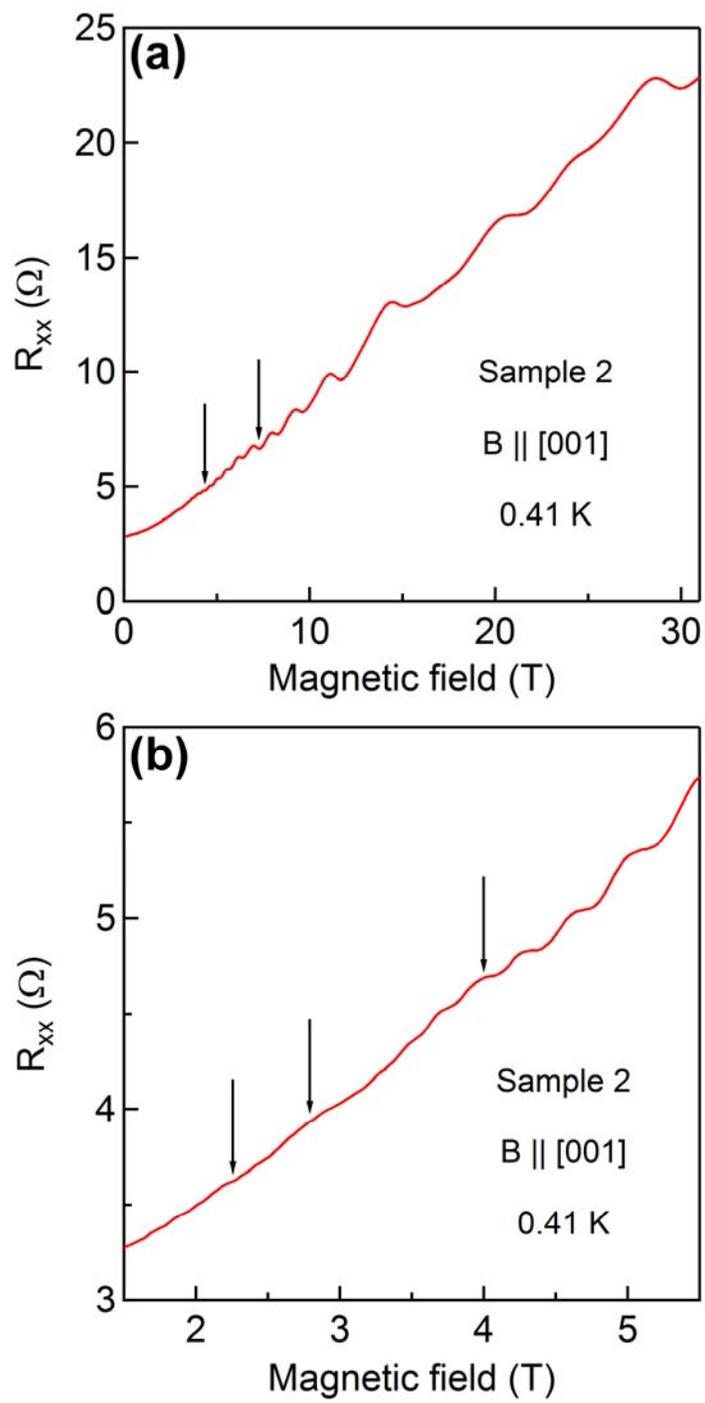

Fig. 5.

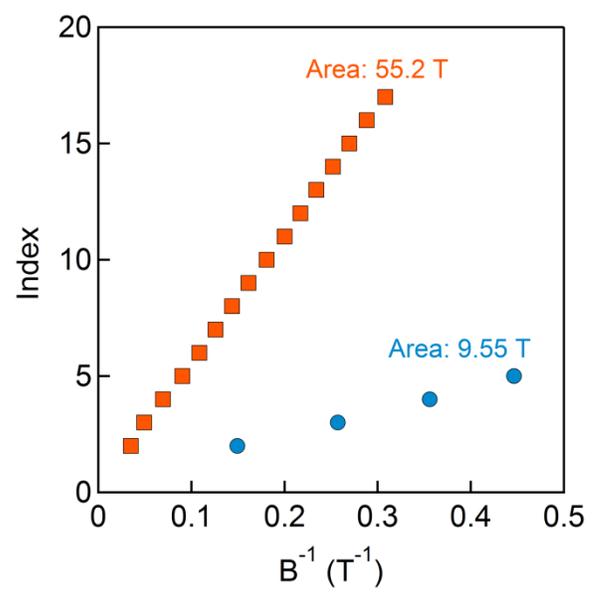

Fig. 6.

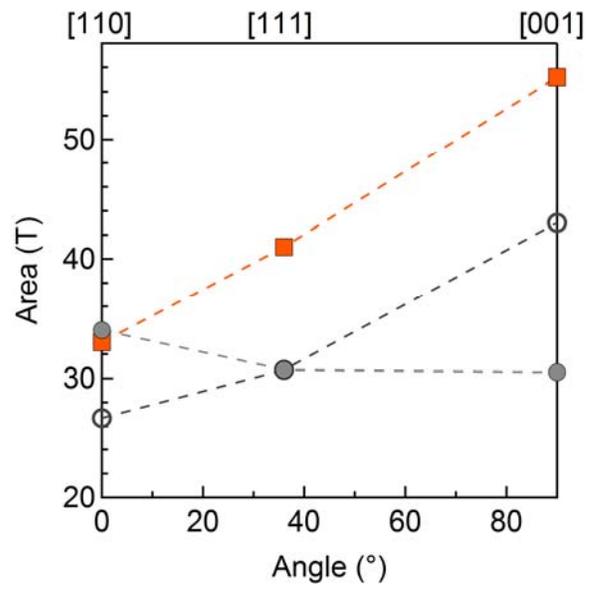

Fig. 7

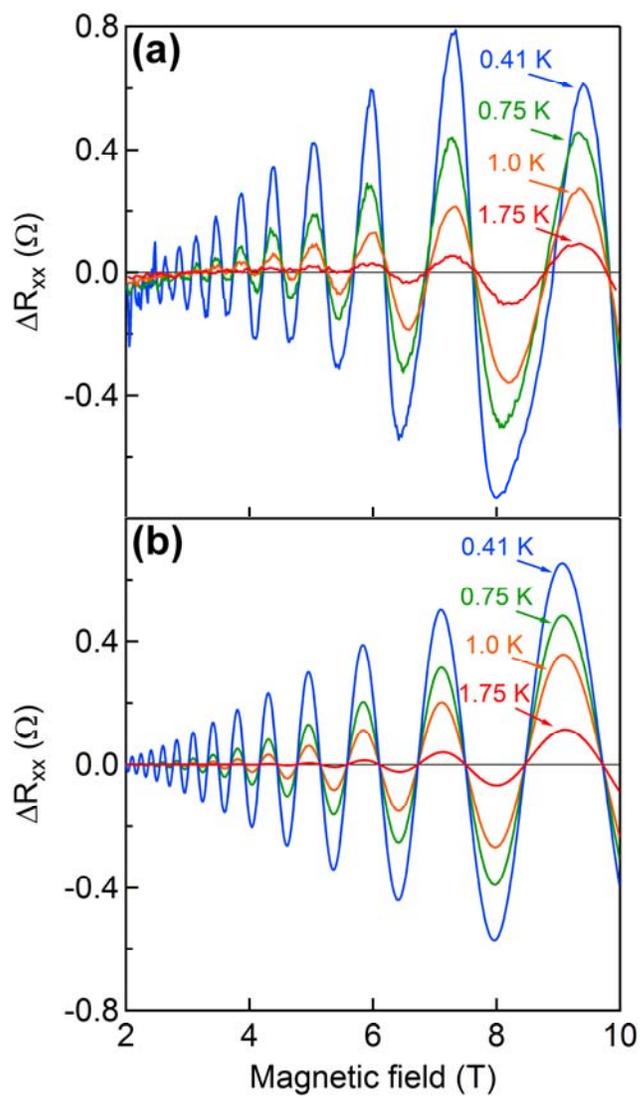

Fig. 8.

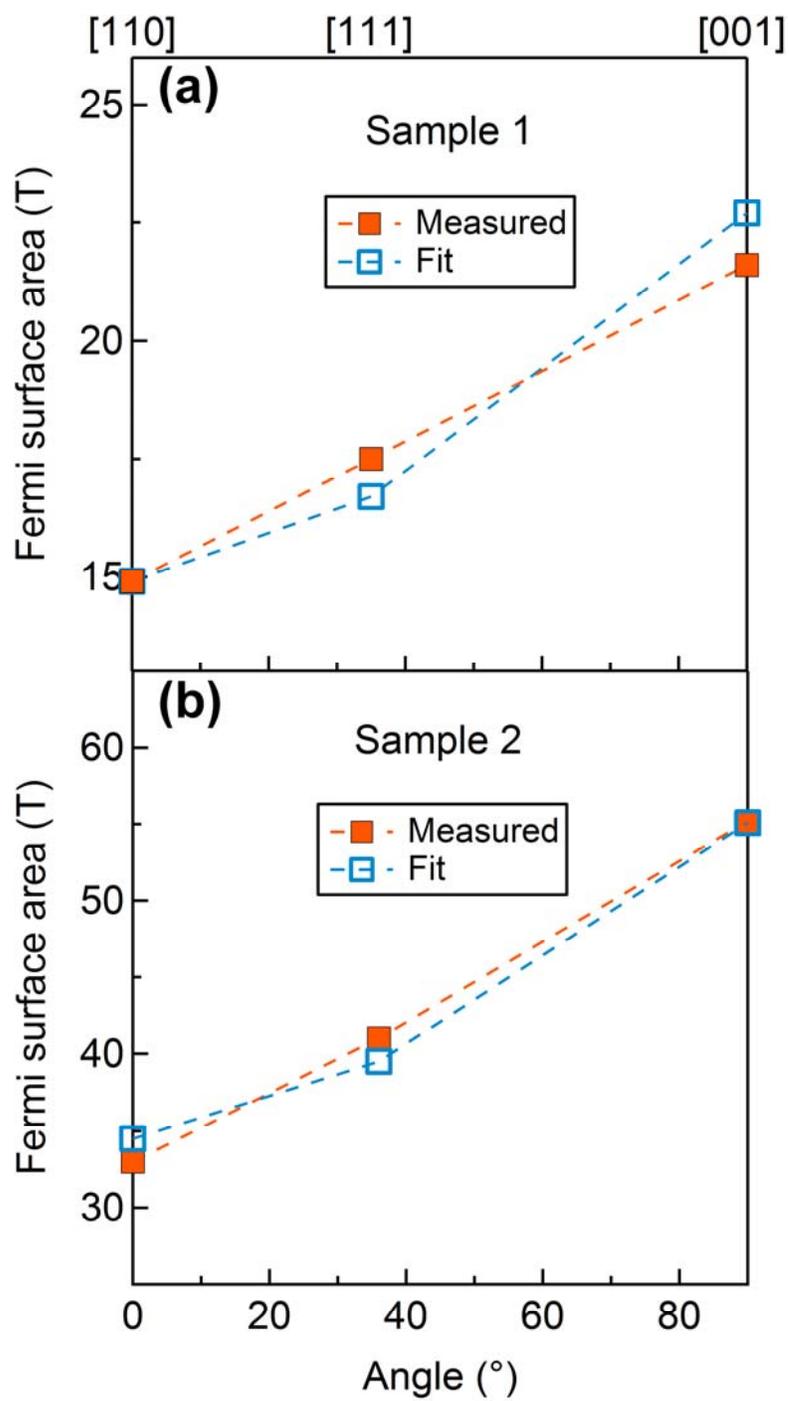